\documentclass[twocolumn]{aastex62}

\usepackage{apjfonts}

\shorttitle{Bow Shock Around FY V\MakeLowercase{ul}}
\shortauthors{Bond et al.}

\newcommand{\Ha}{H$\alpha$}
\newcommand{\Hb}{H$\beta$}
\newcommand{\Hg}{H$\gamma$}

\newcommand{\OIII}{\ion{O}{3}}

\newcommand{\kms}{{\>\rm km\>s^{-1}}}

\def\hei{\ion{He}{1}}

\def\nai{\ion{Na}{1}}

\def\oiii{\ion{O}{3}}

\def\Gaia{{\it Gaia}}

\newcommand{\TESS}{{\it TESS}}

\def\FY{FY~Vul}
\def\St{StDr~90}

\begin{document}

\title{A Bow-Shock Nebula Around the Z~Camelopardalis-type Cataclysmic Variable FY~Vulpeculae\footnote{Based in part on observations obtained with the Hobby-Eberly Telescope (HET), which is a joint project of the University of Texas at Austin, the Pennsylvania State University, Ludwig-Maximillians-Universit\"at M\"unchen, and Georg-August Universit\"at G\"ottingen. The HET is named in honor of its principal benefactors, William P. Hobby and Robert E. Eberly.} }

\author[0000-0003-1377-7145]{Howard E. Bond}
\affil{Department of Astronomy \& Astrophysics, Penn State University, University Park, PA 16802, USA}
\affil{Space Telescope Science Institute, 
3700 San Martin Dr.,
Baltimore, MD 21218, USA}

\author[0009-0008-5193-4053]{Calvin Carter}
\affil{Rocket Girls Ranch Observatory,
7215 Paldao Dr.,
Dallas, TX 75240, USA }

\author[0009-0002-4014-2559]{Eric Coles}
\affil{868 Baker Court. 
Glen Ellyn, IL 60137, USA }
\affil{Sierra Remote Observatories,
42130 Bald Mountain Rd., 
Auberry, CA  93602, USA}


\author[0009-0005-3715-4374]{Peter Goodhew}
\affil{Deep Space Imaging Network, 108 Sutton Court Rd., London, W4 3EQ, UK}



\author[0009-0009-3986-4336]{Jonathan Talbot}
\affil{Stark Bayou Observatory, 1013 Conely Cir., Ocean Springs, MS 39564, USA}


\author[0000-0003-2307-0629]{Gregory R. Zeimann}
\affil{Hobby-Eberly Telescope, University of Texas at Austin, Austin, TX 78712, USA}

\correspondingauthor{Howard E. Bond}
\email{heb11@psu.edu}

\begin{abstract}

We present deep images of the faint nebulosity StDr~90, which we have discovered  surrounds the cataclysmic variable (CV) star FY~Vulpeculae. Archival photometric and spectroscopic observations, and a new optical spectrum, confirm that \FY\ belongs to the Z~Camelopardalis subclass of CVs. Our imagery, obtained by accumulating long exposures with amateur telescopes equipped with CMOS cameras, shows a prominent bow shock in the light of [\oiii] $\lambda$5007, collisionally excited in front of the star as it passes through a relatively dense region in the surrounding interstellar medium (ISM). \FY\ also lies near the edge of an extended faint \Ha-emitting nebula, which we interpret as a ``recombination wake,'' i.e., a Str\"omgren zone recombining after being photoionized by the star's ultraviolet radiation. \FY\ joins five other CVs known to be associated with optical bow shocks and off-center nebulae. All of them are characterized by luminous accretion disks, which drive fast winds into the ISM that produce the bow shocks.

\null\vskip 0.2in

\end{abstract}



\section{Introduction \label{sec:intro} }

\subsection{Faint Nebulae Around Cataclysmic Variable Stars \label{subsec:faint_nebulae} }

This is the fourth in a series of papers that present deep optical images, obtained by amateur astronomers, which reveal faint nebulae around cataclysmic variables (CVs).

For reviews of CVs, see \citet{Warner1995}, \citet{Osaki1996}, \citet{Szkody2012}, and \citet{Sion2023}.
In summary, CVs are close binaries in which a Roche-lobe--filling star transfers mass to a compact companion, usually a white dwarf (WD)\null. In most CVs, the transferred mass forms an accretion disk around the WD; from there the material eventually falls primarily onto the WD, but some of it is ejected from the accretion disk as a fast wind.  There are several subclasses of CVs, defined according to their variability behavior. For our studies the most relevant are: (1)~dwarf novae (DNe), in which the mass-transfer rate is low and the accretion disk is usually optically thin, but becomes optically thick and brighter during occasional outbursts; (2)~nova-like variables (NLVs), where the mass-transfer rate is so high that the accretion disk remains optically thick and bright most or all of the time; and (3)~classical novae (CNe), where the hydrogen accumulated on the surface of the WD ignites nuclear fusion, producing a nova explosion. Following their eruptions, CNe generally behave similarly to NLVs.

In Paper~I \citep{BondSYCnc2024} we reported the discovery of a faint nebula associated with the CV SY~Cancri. This nebulosity had been noticed first in the course of searches  aimed at finding new faint planetary nebulae (PNe; see, for example, \citealt{Jacoby2010}, \citealt{LeDu2022}, and \citealt{Ritter2023}). These investigations, mainly carried out by amateurs, are based on inspection of publicly available wide-field sky-survey images. However, a literature search revealed that this particular nebula surrounds a known CV, SY~Cnc, and thus is not likely to be a classical PN\null. Following up on this serendipitous discovery, several members of our team obtained deep narrow- and broad-band images of the nebula, accumulating long integration times on small telescopes equipped with modern low-noise detectors. This imagery revealed a bow-shock morphology, especially bright in the light of [\oiii] $\lambda$5007, as well as the fact that the CV and bow shock lie off-center within a surrounding large and faint \Ha-emitting nebula. We interpreted the SY~Cnc nebula as the result of a random high-speed encounter between a CV with a fast stellar wind and an interstellar cloud. The off-center \Ha-emitting nebula is ionized material that is recombining following the passage of the ultraviolet-luminous star.

SY~Cnc is classified as a Z~Camelopardalis variable \citep[ZCV; see the review by][]{Simonson2014}. ZCVs are a subset of CVs that share properties of both DNe and NLVs. They exhibit eruptions similar to those of DNe, but can occasionally remain in ``standstills'' at an intermediate brightness level, for intervals of a few days up to many years. The majority of the time, however, SY~Cnc itself exhibits regular outbursts; these occur at intervals of about 3.5 to 4~weeks, with an amplitude of about 2~mag (see Paper~I for details).

Discoveries and follow-up deep imagery of two more cases of faint bow shocks and off-center \Ha\ nebulae around NLVs were presented in our Paper~II \citep{BondLSPeg2025}. These nebulae, surrounding LS~Pegasi and ASASSN-V J205457.73+515731.9, have morphologies remarkably similar to the nebula associated with SY~Cnc. They join two further NLVs that had found earlier to be associated with bow shocks: BZ~Camelopardalis and V341~Arae (see Paper~II for details and references, and Table~\ref{tab:list} at the end of the present paper). 

In Paper~III \citep{BondAntlia2025} we presented the discovery, with follow-up imagery, of yet another faint and previously unknown nebula around a CV\null. This time the star is a little-studied ZCV, ASASSN-19ds, in the southern constellation Antlia. Unlike the objects discussed above, the nebula is approximately centered on the star, and no bow shock is detected.  These findings make the ASASSN-19ds nebula reminiscent of 
the faint nebulosities discovered around Z~Cam itself \citep{Shara2007, SharaZCam2012, Shara2024} and the ZCV AT~Cancri \citep{SharaATCnc2012}. These nebulae are plausibly attributed to ejection from CN outbursts of the central binaries that occurred several centuries to more than a millennium ago. 



\subsection{The Faint Nebula StDr~90 around the Cataclysmic Variable FY~Vulpeculae \label{subsec:discovery_of_nebula} }

For the past several years, H.E.B. and colleagues have been carrying out a spectroscopic survey of central stars of faint PNe. The spectroscopy is obtained with the queue-scheduled (see \citealt{Shetrone2007PASP}) second-generation ``blue'' Low-Resolution Spectrograph (LRS2-B; \citealt{Chonis2016}) on the 10-m Hobby-Eberly Telescope (HET; \citealt{Ramsey1998,Hill2021}), located at McDonald Observatory in west Texas, USA\null. Details of the spectrograph and data reduction are given below (Section~\ref{subsec:spectroscopy}), as well as in a series of papers resulting from the PN survey. The two most recent of these are \citet{BondAbell572024} and \citet{Reindl2024}.

A major source of newly discovered and faint PNe, whose nuclei are targeted in this survey, is the PlanetaryNebulae.net website,\footnote{\url{https://planetarynebulae.net/EN/index.php}} maintained by amateurs located primarily in France. In 2025 July, in the course of inspecting a list of candidate PNe identified by X.~Strottner and M.~Drechsler, H.E.B. noticed that the nebula StDr~90 appears to have a relatively bright (14th-mag), blue central star. This object proved to be a known CV, FY~Vulpeculae. Strottner and Drechsler found StDr~90 on images from the Isaac Newton Telescope Photometric H-Alpha Survey (IPHAS; \citealt{Drew2005}), and they reported the nebula to have a diameter of $2\farcm5$. StDr~90 is only marginally visible on photographic imagery from the Space Telescope Science Institute Digitized Sky Survey.\footnote{\url{https://archive.stsci.edu/cgi-bin/dss_form}} For further information about this little-studied nebula, see its listing in the online Hong-Kong/AAO/Strasbourg/H$\alpha$ Planetary Nebulae (HASH) database\footnote{\url{http://hashpn.space/}} \citep{Parker2016, Bojicic2017}. 

The presence of a CV inside the nebula indicated that StDr~90 is likely not a PN, but is instead probably similar to the objects discussed above. H.E.B. communicated the discovery to the amateur co-authors of this paper, who immediately responded by collecting the deep imagery that is presented and discussed below.

\section{FY V\MakeLowercase{ulpeculae}}

\subsection{Discovery and Classification}

Variability of FY~Vul was discovered in 1940 at the Sternwarte Sonneberg by C.~Hoffmeister, and the star was given a designation of AN~122.1940. A finding chart was published some years later \citep{Hoffmeister1957}. Based on the Sonneberg photographic data, \FY\ was classified as a DN by \citet{Richter1961}. This was refined to a ZCV classification by \citet[][where the object is designated CSV~4779]{Meinunger1965}, because of standstills in its light curve. This conclusion was questioned by \citet{Simonson2014}, who investigated data obtained by members of the American Association of Variable Star Observers, and saw no standstills. However, subsequent work (see below) has reconfirmed its membership in the ZCV class, although the standstills occur only rarely. 

Table~\ref{tab:DR3data} lists astrometry and photometry of \FY\ from \Gaia\/ Data Release~3\footnote{\url{https://vizier.cds.unistra.fr/viz-bin/VizieR-3?-source=I/355/gaiadr3}} (DR3; \citealt{Gaia2016, Gaia2023}). The bottom row gives the star's nominal absolute magnitude in the \Gaia\/ system, based on the photometry and the star's trigonometric parallax.\footnote{A Bayesian analysis of {\it Gaia\/} data by \citet{BailerJones2021} gives a distance of $573.0^{+6.7}_{-5.2}$~pc.} Here we have adopted an interstellar extinction of $A_G=0.32$~mag at the distance of the star, obtained using the online {\tt GALExtin} tool\footnote{\citet{Amores2021}; \url{http://www.galextin.org/}} and based on the 3D extinction map of \citet{Chen2019}.

\begin{deluxetable}{lc}
\tablecaption{\Gaia\/ DR3 Data for FY Vul \label{tab:DR3data} }
\tablehead{
\colhead{Parameter}
&\colhead{Value}
}
\decimals
\startdata
RA (J2000)  & 19 41 39.933  \\
Dec (J2000) &  +21 45 58.36  \\
$l$ [deg] &  57.90  \\
$b$ [deg] &  $-0.61$ \\
Parallax [mas] & $1.7078  \pm 0.0179 $   \\
$\mu_\alpha$ [mas\,yr$^{-1}$] & $-7.417  \pm 0.018 $  \\
$\mu_\delta$ [mas\,yr$^{-1}$] & $ -18.827 \pm 0.018 $   \\
$G$ [mag] & 14.38  \\
$G_{\rm BP}-G_{\rm RP}$ [mag] & 0.47  \\
$M_G$ [mag] & +5.3  \\
\enddata
\end{deluxetable}

\FY\ has been the subject of only a few studies. In this paper we are primarily concerned with its surrounding nebula. However, in the following subsections we briefly review the main photometric and spectroscopic properties of the star. The cited papers should be consulted for further information about \FY\ itself. 

\subsection{Long-Term Light Variations}

Photometric variability of \FY\ was studied by \citet{Kato2019}. He investigated data obtained by the sky patrol of the All-Sky Automated Survey for SuperNovae \citep[ASAS-SN;][]{Kochanek2017}, which measures optical magnitudes in the $V$ or $g$ bands at intervals generally of a few days. Based on the ASAS-SN observations, Kato placed \FY\ into a subclass of ZCVs called ``IW~And variables,'' defined as ZCVs that increase in brightness after a standstill (see, for example, \citealt{Szkody2013}). This behavior is opposite to that of most ZCVs, which decline at the ends of their standstills.

The IW~And phenomenon was discussed in more detail by \citet{Kimura2020}, and an updated ASAS-SN light curve of \FY\ was included. It shows that the star is often in a mode where it shows repeated outbursts on a timescale of roughly 35 days, but at other times it drops into a state lacking such eruptions. 

\begin{figure}[h]
\centering
\includegraphics[width=0.47\textwidth]{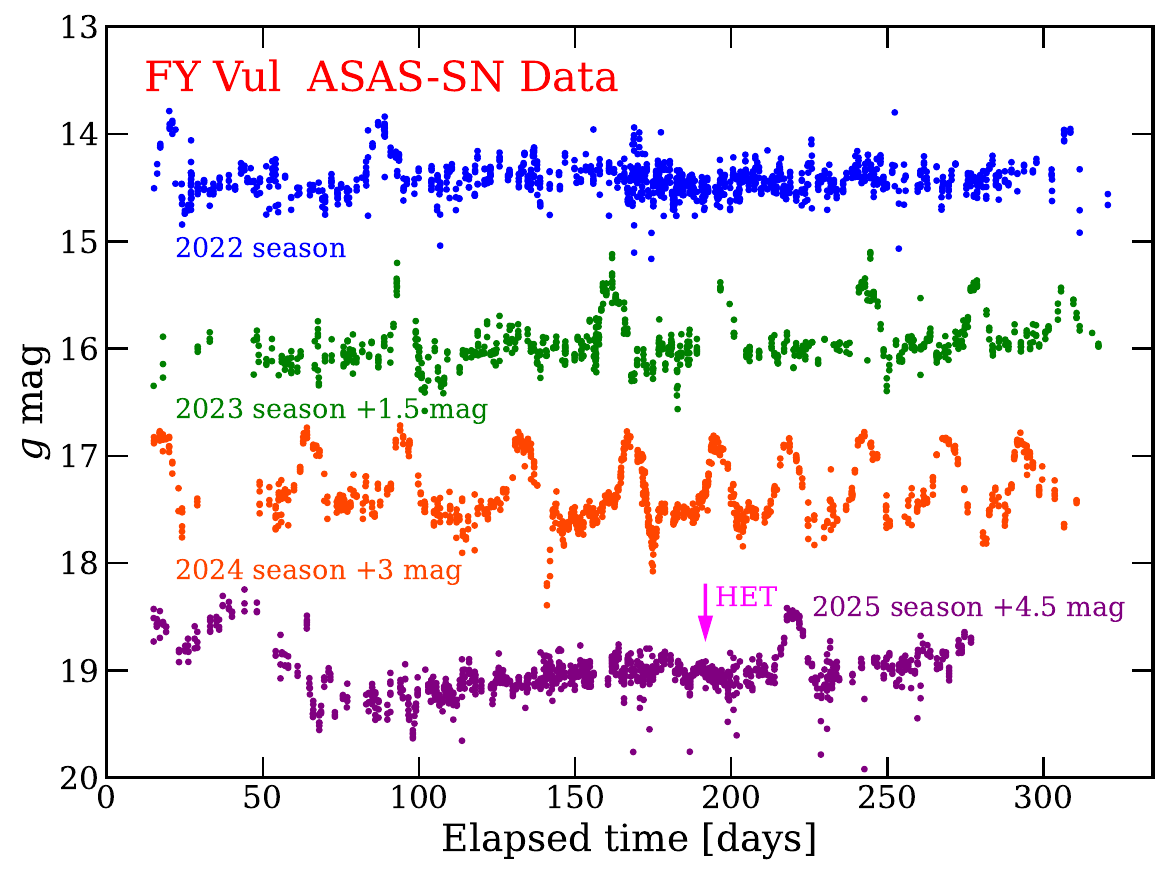}
\caption{
Seasonal $g$-band light curves of \FY\ based on data in the ASAS-SN archive. Blue points are for 2022 February~9 to December~12; green points (offset by +1.5~mag) are for 2023 February~12 to December~12; orange points (offset by +3~mag) are for 2024 February~17 to December~9; and purple points (offset by +4.5~mag) are for 2025 February~3 to October~23. The star sometimes shows a ``sawtooth'' light curve, typical of Z~Cam variables, with a variable time interval between successive brightness maxima; but at other times these outbursts subside, sometimes for extended intervals. The magenta arrow above the 2025 light curve marks the date of the HET spectrum shown below in Figure~\ref{fig:hetspectrum}.
\label{fig:asassn_seasons}
}
\end{figure}

We investigated the most recent photometric behavior of \FY\ by downloading\footnote{from \url{https://asas-sn.osu.edu}}  the ASAS-SN data for the past four yearly seasons, 2022 to the present. These data are plotted in Figure~\ref{fig:asassn_seasons}. Here the seasonal light curves are shifted down successively by 1.5~mag from year to year. During the 2022 season, the star showed a few outbursts, but was fairly constant most of the time, apart from small-amplitude flickering. About a half-dozen outbursts were seen in 2023, and these became regular events in 2024, recurring at intervals ranging from $\sim$23 to 39~days. ``Sawtooth'' variations like this are commonly seen in ZCVs (see, for example, the light curves displayed in \citealt{Shafter2005}). On a few occasions in 2024, sharp transient dips in brightness occurred immediately after the outbursts, a behavior occurring in several other IW~And-type ZCVs \citep[e.g.,][]{Kato2019}. { In 2025, the outbursts largely disappeared, and a new behavior appeared. The overall brightness level was seen to fade to lower than it had been in the three previous years---an apparent standstill---and then the mean brightness started a slow increase. This increase continued to the end of the plotted data, interrupted by a single outburst of the type seen in previous years.
}

\subsection{Time-Series Photometry \label{subsec:time_series} }

Turning to high-cadence photometry, \citet{Bruch2024} investigated \FY\ using data from the {\it Transiting Exoplanet Survey Satellite\/} (\TESS; \citealt{Ricker2015}) mission. He calculated power spectra for the continuous measurements collected in four \TESS\/ ``Sectors'' with durations of $\sim$27~days each, obtained in 2019, 2021 (two Sectors), and 2022. Two dominant photometric periods were seen, one with a period of 0.2014~days (4.83~hr, present during two of the four Sectors), and the other having a shorter period of 0.1910~days (4.58~hr, present during three Sectors). During one Sector, both periods were detected in the power spectrum. 

To our knowledge, the orbital period of \FY\ is unknown. Bruch suggests, however, that the longer photometric period corresponds to the orbital period, and that the period shorter by $\sim$5\% is that of a ``negative superhump.'' This phenomenon is seen in the light curves of SY~Cnc (Paper~I), and of LS~Peg (Paper~II), where the orbital periods are known. The superhump periods are shorter than them by 5\% and 2\%, respectively. See our previous papers for references to astrophysical interpretations of negative superhumps.  

\begin{figure}[h]
\centering
\includegraphics[width=0.47\textwidth]{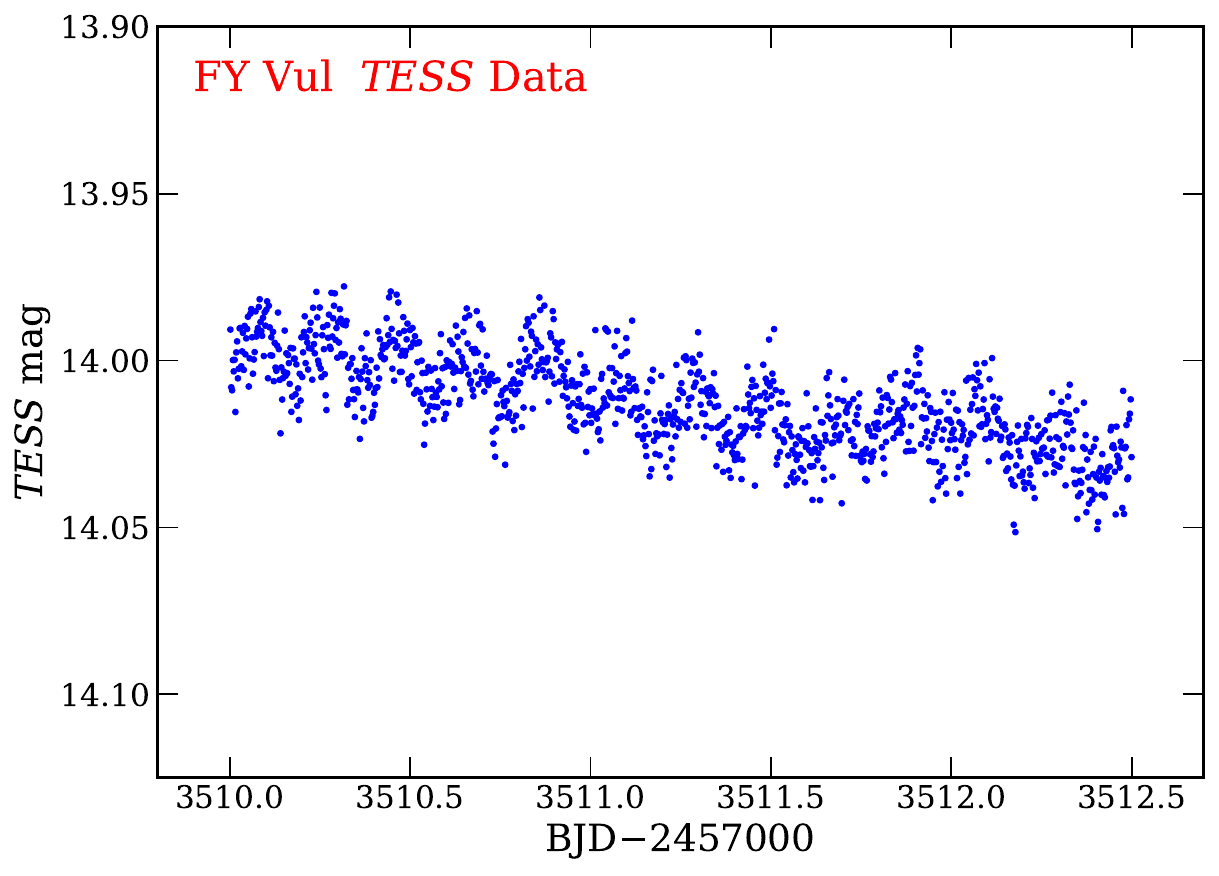}
\caption{
A representative \TESS\/ light curve for \FY, covering a 2.5-day interval in 2024 July at a 200~s cadence. A low-amplitude variation with a period of 0.2014~days is seen, possibly corresponding to the orbital period of the binary (see text).
\label{fig:tess}
}
\end{figure}

\FY\ was observed again by \TESS\/ in 2024 July, during the Sector~81 run. We obtained its light curve using the online {\tt TESSExtractor} tool.\footnote{\citet{Serna2021}; \url{https://www.tessextractor.app}} A representative 2.5-day segment of this light curve is shown in Figure~\ref{fig:tess}. Variations with a peak-to-peak amplitude of $\sim$0.03~mag\footnote{Note that, because of the low spatial resolution of \TESS, the image of \FY\ is blended with several neighboring stars, and the true amplitude of the variations is likely higher.} are seen, with a period corresponding to the longer period detected in the earlier \TESS\/ data by \citet{Bruch2024}.

\subsection{Spectroscopy \label{subsec:spectroscopy} }

To our knowledge, the only published (until now) spectroscopy of \FY\ was presented by \citet{Downes1995}. Their spectrum, obtained in 1992, showed a very blue continuum with \Ha\ in emission. There were broad weak absorption features at \hei\ $\lambda$5875, \Hb, and \Hg, with superposed faint emission cores.\footnote{An unpublished spectrum obtained in the late 1970s in a spectroscopic survey of CVs described by \citet{Bond1979} was, although noisy, similar to that described by \citet{Downes1995}.} Such a spectrum is typical of a CV near maximum light.

Inspired by our recognition of its surrounding nebulosity, we obtained a spectrum of \FY\ on 2025 July~30, using the LRS2-B spectrogram on HET (see Section~\ref{subsec:discovery_of_nebula}). 
Briefly, LRS2-B employs a dichroic beamsplitter to send light simultaneously into two units: the ``UV'' channel (covering 3640--4645~\AA\ at a resolving power of 1910), and the ``Orange'' channel (covering 4635--6950~\AA\ at a resolving power of 1140). Data reduction is carried out by co-author G.R.Z., using the \texttt{Panacea}\footnote{\url{https://github.com/grzeimann/Panacea}} and \texttt{LRS2Multi}\footnote{\url{https://github.com/grzeimann/LRS2Multi}} packages.
Further details of the LRS2-B spectrograph and data-reduction procedures are given, for example, in \citet{BondPaperI2023}.

\begin{figure}
\centering
\includegraphics[width=0.47\textwidth]{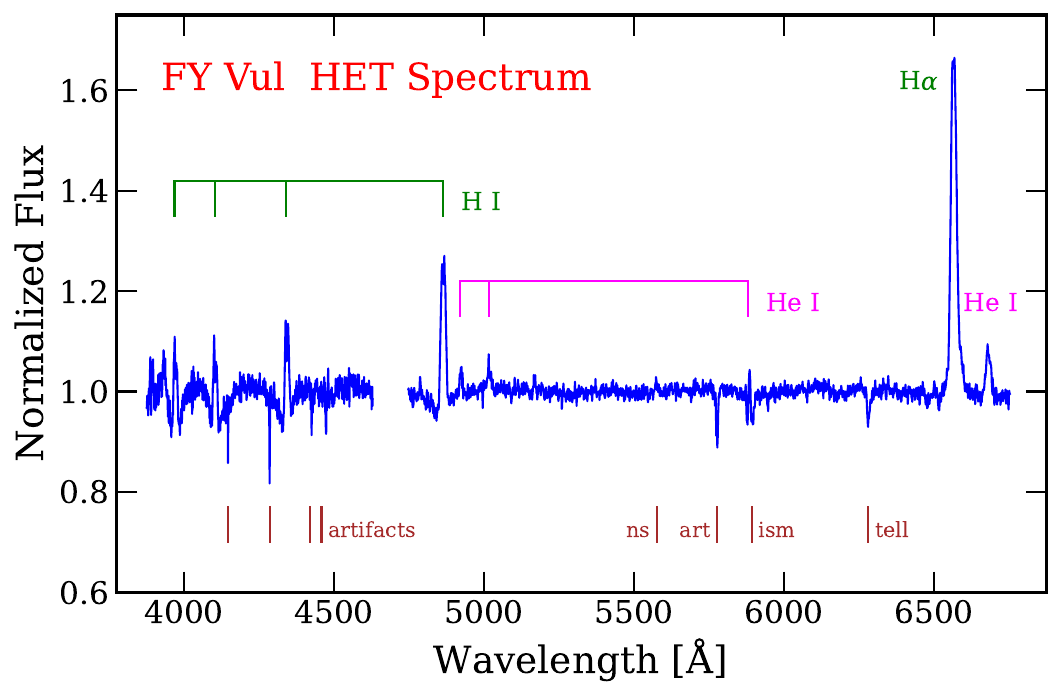}
\caption{
Normalized HET LRS2 spectrum of FY~Vul on 2025 July~30, showing emission lines of hydrogen and \hei. Higher members of the Balmer series, and \hei\ $\lambda$5875, exhibit surrounding shallow broad absorption wings. Several features due to instrument artifacts (``artifacts'' and ``art''), night-sky emission (``ns''), and interstellar \nai\ and telluric absorption (``ism'' and ``tell'') are marked. 
\label{fig:hetspectrum}
}
\end{figure}

Our spectrum is plotted in Figure~\ref{fig:hetspectrum}, where it is normalized to a flat continuum. The Balmer series is seen in emission, superposed on broad absorption at \Hb\ and higher members of the series. Several lines of \hei\ are also in emission, with $\lambda$5875 superposed on broad absorption wings. Here the Balmer and \hei\ emission lines are much stronger than in the 1992 spectrum of \citet{Downes1995}. This enhancement is consistent with the appearance of a ZCV's spectrum near minimum light. A magenta arrow in Figure~\ref{fig:asassn_seasons} marks the date of the LRS2-B spectrum, which indeed was obtained at a time when the star was near its baseline magnitude. The spectrum is very similar to those we obtained for the NLV ASASSN-V J205457.73+515731.9, presented in Paper~II.

To summarize the photometric and spectroscopic material, \FY\ is well established as a typical ZCV, belonging to the IW~And subclass. Beyond this, the available information, unfortunately, is fairly limited. In particular, the orbital period proposed by \citet{Bruch2024} lacks definitive confirmation.

\section{Deep Imaging \label{sec:deep_imaging} }

\subsection{Observations}

We imaged StDr~90/\FY\ and its surrounding field using six different telescopes, located at sites in California, Mississippi, and Texas in the United States, and in Spain. Exposures were obtained between 2025 July~22 and August~3. Telescope apertures ranged from 6 to 20~inches. Details of these  instruments, including the optical filters and CMOS cameras, are given in Paper~II\null. Long exposures were accumulated in \Ha\ and [\oiii] $\lambda$5007 filters in order to image the very faint nebulosity, and shorter exposures were taken in red, green, and blue filters to show the stellar field. Table~\ref{tab:fyvul_exposures} lists the exposure times. The telescope numberings in column~1 are the same as in Paper~II, Table~4. Grand total exposure time was 174.8~hr.

\begin{deluxetable*}{lccccccc}[h]
\tablecaption{Exposure Times [s] on 
StDr~90/FY Vul\label{tab:fyvul_exposures} }
\tablehead{
\colhead{Telescope\tablenotemark{a}}
&\colhead{Aperture}
&\colhead{Observer(s)}
&\colhead{\Ha}
&\colhead{[\oiii]}
&\colhead{R}
&\colhead{G}
&\colhead{B}
}
\startdata
1    & 20 in & Coles & $47\times600$ & $69\times600$ & $22\times600$ & $23\times600$ & $35\times600$ \\
2, 3 & Twin 6 in & Goodhew & $488\times300$ & $339\times300$ & $29\times300$ & $29\times300$ & $29\times300$ \\
4    & 14 in & Goodhew & $192\times300$ & $207\times300$ & \dots & \dots & \dots \\
5    & 6 in & Talbot & $59\times1200$ & \dots & \dots & \dots & \dots \\
6    & 13.8 in & Carter \& Talbot & $79\times480$ & \dots & $54\times60$ & $50\times60$ & $50\times60$ \\
Total exp.\ [hr] & & & 94.70 & 57.00 & 6.98 & 7.08 & 9.08 \\
\enddata
\tablenotetext{a}{Technical details for the telescopes and instrumentation are given in Table~4 in Paper~II.}
\end{deluxetable*}

\subsection{Imagery and Interpretation \label{subsec:imagery} }

Pre-processing of the frames was done using {\tt CCDStack},\footnote{\url{https://ccdware.com/ccdstack_overview}} {\tt PixInsight},\footnote{\url{https://pixinsight.com}} and {\tt Photoshop}.\footnote{\url{https://www.adobe.com/products/photoshop.html}} The large number of frames was combined using {\tt AstroPixelProcessor}.\footnote{\url{https://www.astropixelprocessor.com}}

\begin{figure*}
\centering
\includegraphics[width=6in]{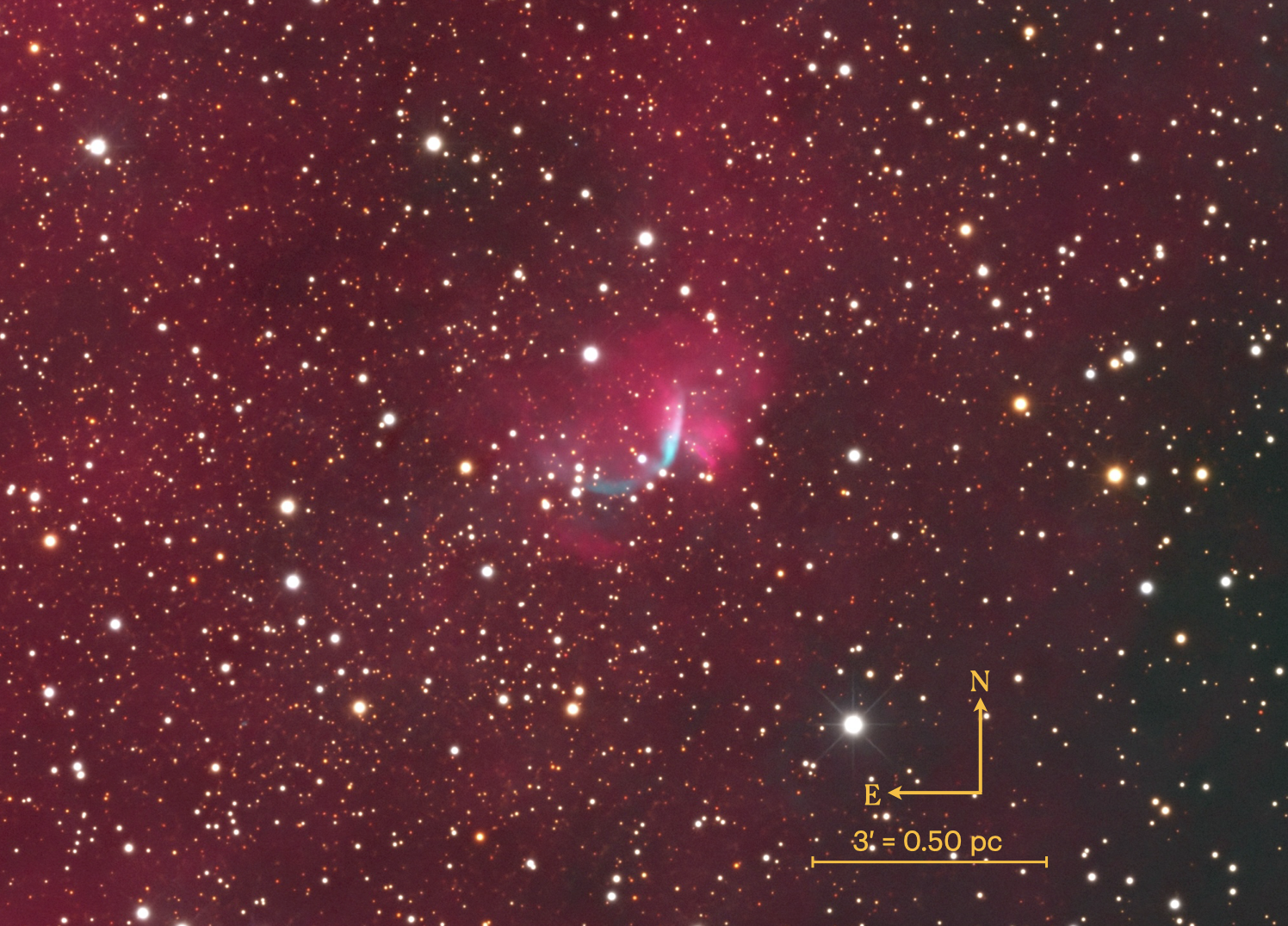}
\caption{
Color image of the nebula StDr~90 around the cataclysmic variable \FY, created from 174.8~hours of exposure time in RGB, \Ha, and [\oiii] $\lambda$5007 filters, as described in the text. \Ha\ is assigned to the red channel, and [\OIII]  to the green and blue channels. Orientation and scale of the image (including a conversion to linear size at the distance of the star) are indicated at the lower right. \FY\ is identified in Figure~\ref{fig:fyvul_zoom}.
\label{fig:fyvul_deepimage}
}
\end{figure*}

Figure~\ref{fig:fyvul_deepimage} displays a color picture of the nebula, in a rendition created by combining all of the frames listed in Table~\ref{tab:fyvul_exposures}. For this presentation, an ``HOO'' palette was employed, with \Ha\ assigned to the red channel, and [\OIII] $\lambda$5007 to the green and blue channels. The orientation and angular scale are indicated at the lower right, along with the conversion to a linear scale at the known distance of \FY.

Figure~\ref{fig:fyvul_deepimage} shows that the low-Galactic-latitude field surrounding \St\ is overlain with extensive faint \Ha\ emission (except in a region of high extinction to the west and southwest). The \St\ nebula itself has the appearance of being a location of enhanced \Ha\ surface brightness, covering a region with a physical scale of about 0.5~pc, lying within this network of faint emission (although, of course, much of it may be at different distances than \FY). 

\begin{figure*}
\centering
\includegraphics[width=5in]{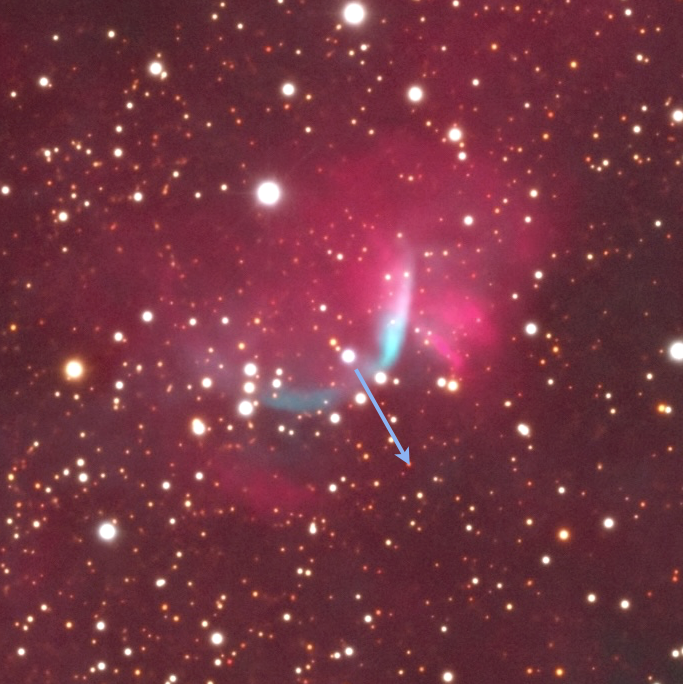}
\caption{Close-up of the StDr~90 nebula. The arrow that extends from the position of \FY\ is oriented in the direction of the star's proper motion relative to the Galactic-rotation standard of rest at the star's location.
\label{fig:fyvul_zoom}
}
\end{figure*}

In Figure~\ref{fig:fyvul_zoom} we zoom in on the image of the StDr~90/\FY\ nebula, in order to show its morphology more clearly. Superposed is a blue arrow with its base located at \FY\ itself, and oriented to show the direction of the star's proper motion. Here the position angle has been adjusted by $+7\fdg9$, relative to the absolute \Gaia\/ value of $201\fdg5$, in order to correct for the effect of differential Galactic rotation.\footnote{For details of this correction, see our discussion in Paper~II\null. To calculate the correction, we used a {\tt python} code created by S.~del Palacio, along with the Oort constants, referenced in Paper~II.} The direction shown is thus made relative to the local standard of rest (LSR) at the distance of the star. The transverse space velocity of \FY, relative to this LSR, is $50.6\pm0.5\,\kms$.

Figures~\ref{fig:fyvul_deepimage} and~\ref{fig:fyvul_zoom} clearly show the morphology of a bow shock, in the form of a bright parabolic rim to the southwest of \FY. 
As discussed in our previous papers, and as is well known, a bow-shock morphology is the signature of a fast wind from the star, colliding with the interstellar medium (ISM) as the star passes through it at high velocity. The arrow in Figure~\ref{fig:fyvul_zoom} shows that the motion of the star is in a direction consistent with the interpretation as a bow shock.

\begin{figure}
\centering
\includegraphics[width=0.47\textwidth]{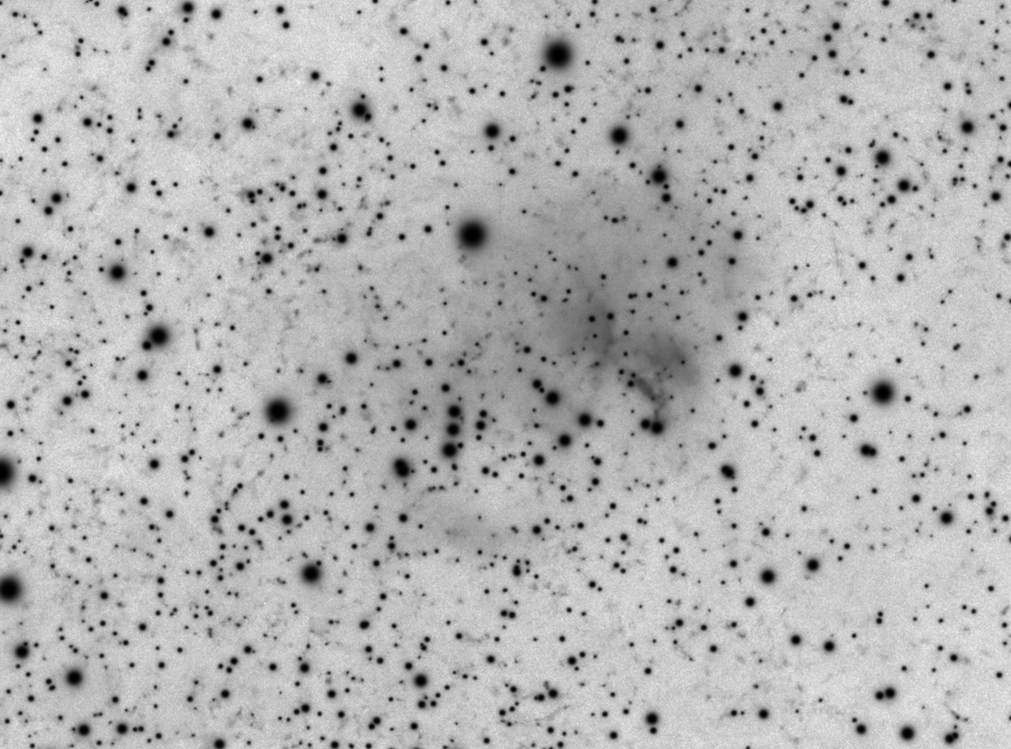}
\includegraphics[width=0.47\textwidth]{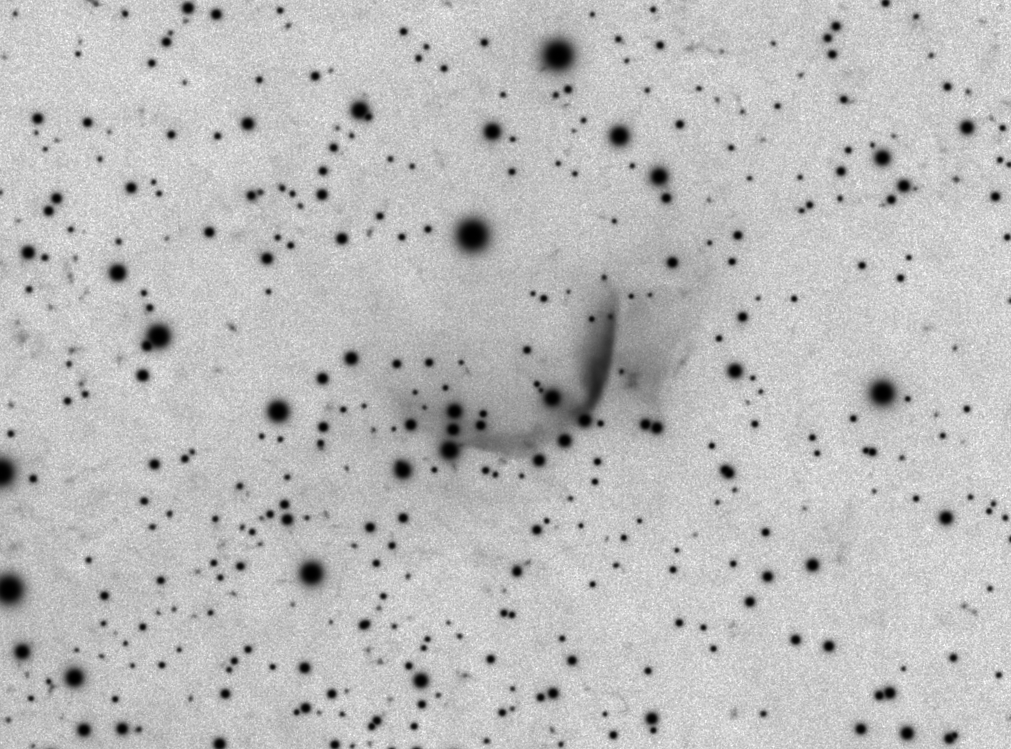}
\caption{Narrow-band images of the \St\ nebula in \Ha\ (top) and [\oiii] $\lambda$5007 (bottom).
\label{fig:BandWframes}
}
\end{figure}

The two panels in Figure~\ref{fig:BandWframes} show narrow-band images of the nebula separately in \Ha\ (top) and [\oiii] (bottom). This imagery shows that the bow shock  is prominent in the light of [\oiii], but is very faint in \Ha. Our interpretation is that the [\oiii] emission is collisionally excited as the stellar wind plows into the ISM\null. The surrounding \Ha\ emission nebula is primarily photoionized by UV and X-ray emission from the star. This material has a patchy distribution, with the highest density lying to the northwest of the star. A more subtle detail is that the [\oiii] emission along the bow shock is brightest to the west and northwest of the star, which is where the \Ha-emitting material is densest. Note that the bow shock is not symmetric around the star, but is bent back on the northwest side, likely due to the stellar wind colliding with the densest ISM material at this location.

\FY\ is located well off-center with respect to the \Ha\ nebulosity, which lies mostly behind the path of the star's proper motion as well as to the north and northwest. These off-center \Ha-emitting nebulae are seen in all of the CVs associated with bow shocks, as discussed below and in Papers~I and II; they are likely to be ionized material that is recombining following passage of the star.  



\section{Summary and Future Work}

In this paper we present deep imaging of the nebula \St, which surrounds the Z~Cam variable star \FY. The imagery was obtained by advanced amateur astronomers, using small telescopes and accumulating extremely long exposure times. The images reveal that \FY\ is associated with a bow shock, bright in the light of collisionally excited [\oiii]. The bow shock is located in front of the star, in the direction of its supersonic motion with respect to the relatively stationary ISM at its location. The star also lies near the front edge of a faint \Ha-emitting nebula.


\FY\ thus joins the five ZCVs and NLVs discussed in our Paper~II and papers cited therein; all of the nebulae around the stars show bow shocks,  and the stars lie off-center in faint \Ha-emitting nebulae.  In Table~\ref{tab:list} we update a table presented in Paper~II that lists these objects, by adding \FY. Objects are listed in order of discovery dates for the nebulae. The nomenclature used in column~1 for four of the nebulae is that of the HASH PN catalog (see Section~\ref{subsec:discovery_of_nebula}), although none of the nebulae are actually classical PNe.


\begin{deluxetable*}{lccccccc}
\tablecaption{Nova-like Variables Associated with Bow Shocks and Off-Center \Ha\ Nebulae \label{tab:list} }
\tablehead{
\colhead{Nebula}
&\colhead{Star}
&\colhead{Distance}
&\colhead{$P_{\rm orb}$}
&\colhead{$G$}
&\colhead{$E(B-V)$}
&\colhead{$M_G$}
&\colhead{References\tablenotemark{a}}\\
\colhead{ }
&\colhead{ }
&\colhead{[kpc]}
&\colhead{[hr]}
&\colhead{[mag]}
&\colhead{[mag]}
&\colhead{[mag]}
&\colhead{}
}
\decimals
\startdata
EGB 4   & BZ Cam    & 0.37 & 3.69  & 12.97 & 0.11 & +4.8 & (1,2) \\ 
Fr 2-11 & V341 Ara  & 0.16 & 3.65  & 10.81 & 0.02 & +4.8 & (3,4) \\ 
PaEl 1  & SY Cnc    & 0.41 & 9.18  & 12.69 & 0.04 & +4.5 & (5,6) \\ 
\dots   & ASASJ2054\tablenotemark{b} & 0.55 & \dots & 13.67 & 0.11 & +4.6 & (7,8) \\ 
\dots   & LS Peg    & 0.29 & 4.19  & 11.89 & 0.03 & +4.5 & (9,10) \\ 
StDr 90 & FY Vul    & 0.57 & 4.83  & 14.38 & 0.12 & +5.3 & (11,12) \\ 
\enddata
\tablenotetext{a}{First reference for each star is for discovery of the nebula, and the second is for determination of the orbital period: (1)~\citet{Ellis1984}; (2)~\citet{Patterson1996}; (3)~\citet{Frew2006}; (4)~\citet{BondV341Ara2018, CastroSegura2021}; (5)~Paper~I; (6)~\citet{Casares2009}; (7)~\citet{BondASASSN2020}; (8)~Period unknown; (9)~Paper~II; (10)~\citet{Taylor1999}; (11) X.~Strottner \& M.~Drechsler; see Section~\ref{subsec:discovery_of_nebula}; (12) Listed period is photometric and a possible orbital period; see Section~\ref{subsec:time_series}.
} 
\tablenotetext{b}{Full designation is ASASSN-V J205457.73+515731.9; see Paper~II.}
\end{deluxetable*}

As the table shows, all six CVs have similar absolute magnitudes, lying in the narrow range $+5.3>M_G>+4.5$. These relatively high luminosities (among CVs) indicate the presence of luminous accretion disks around the systems' WD components, due to high mass-transfer rates from their donor stars. These accretion disks launch fast winds into the surrounding space. This situation provides especially favorable circumstances for the formation of bow shocks, in cases where the star happens to be passing at high speed through a relatively stationary interstellar cloud. We interpret the faint off-center \Ha\ nebulae as ``recombination wakes,'' i.e.,  Str\"omgren zones that are recombining behind the paths of the stars after being photoionized by their ultraviolet and X-ray radiation.  

As we discussed in Paper~II, we hope our findings encourage 
deep imaging of larger samples of NLVs. Amateurs are equipped to reach deeper surface-brightness levels than many existing surveys, as shown in our series of papers, and they have access to large amounts of observing time.

We urge spectroscopic monitoring of \FY\ with large telescopes, in order to determine definitively its orbital period. As shown in Table~\ref{tab:list}, the other five ZCVs associated with bow shocks all have periods of 3.6 to 4.2~hr (except for the longer 9.2~hr period of SY~Cnc, due to its donor star being a G-type star, rather than the usual low-mass main-sequence donor). The orbital period of \FY\ is likely to be 4.83~hr, based on photometric observations, but this has not been proven directly.


\acknowledgments



The Digitized Sky Surveys were produced at the Space Telescope Science Institute under U.S. Government grant NAG W-2166. The images of these surveys are based on photographic data obtained using the Oschin Schmidt Telescope on Palomar Mountain and the UK Schmidt Telescope. The plates were processed into the present compressed digital form with the permission of these institutions. 


This work has made use of data from the European Space Agency (ESA) mission
{\it Gaia\/} (\url{https://www.cosmos.esa.int/gaia}), processed by the {\it Gaia\/}
Data Processing and Analysis Consortium (DPAC,
\url{https://www.cosmos.esa.int/web/gaia/dpac/consortium}). Funding for the DPAC
has been provided by national institutions, in particular the institutions
participating in the {\it Gaia\/} Multilateral Agreement.


Funding for the \TESS\/ mission is provided by NASA's Science Mission directorate.


This research has made use of the SIMBAD and Vizier databases, operated at CDS, Strasbourg, France.

The Low Resolution Spectrograph~2 (LRS2) on the Hobby-Eberly Telescope was developed and funded by the University of Texas at Austin McDonald Observatory and Department of Astronomy, and by Pennsylvania State University. We thank the Leibniz-Institut f\"ur Astrophysik Potsdam (AIP) and the Institut f\"ur Astrophysik G\"ottingen (IAG) for their contributions to the construction of the integral field units.

We acknowledge the Texas Advanced Computing Center (TACC) at The University of Texas at Austin for providing high-performance computing, visualization, and storage resources that have contributed to the results reported within this paper.

We thank S.~del Palacio, author of the {\tt python} code used in Section~\ref{subsec:imagery}, for pointing out to us the importance of correcting proper motions for the effect of differential Galactic rotation, and for kindly assisting us in implementing his code.




\bibliography{PNNisurvey_refs}

@ARTICLE{BailerJones2021,
       author = {{Bailer-Jones}, C.~A.~L. and {Rybizki}, J. and {Fouesneau}, M. and {Demleitner}, M. and {Andrae}, R.},
        title = "{Estimating Distances from Parallaxes. V. Geometric and Photogeometric Distances to 1.47 Billion Stars in Gaia Early Data Release 3}",
      journal = {\aj},
         year = 2021,
        month = mar,
       volume = {161},
       number = {3},
          eid = {147},
        pages = {147},
          doi = {10.3847/1538-3881/abd806},
archivePrefix = {arXiv},
       eprint = {2012.05220},
 primaryClass = {astro-ph.SR},
       adsurl = {https://ui.adsabs.harvard.edu/abs/2021AJ....161..147B}
}

@ARTICLE{Amores2021,
       author = {{Am{\^o}res}, Eduardo B. and {Jesus}, Ricardo M. and {Moitinho}, Andr{\'e} and {Arsenijevic}, Vladan and {Levenhagen}, Ronaldo S. and {Marshall}, Douglas J. and {Kerber}, Leandro O. and {K{\"u}nzel}, Roseli and {Moura}, Rodrigo A.},
        title = "{GALExtin: an alternative online tool to determine the interstellar extinction in the Milky Way}",
      journal = {\mnras},
         year = 2021,
        month = dec,
       volume = {508},
       number = {2},
        pages = {1788-1797},
          doi = {10.1093/mnras/stab2248},
archivePrefix = {arXiv},
       eprint = {2108.00561},
 primaryClass = {astro-ph.GA},
       adsurl = {https://ui.adsabs.harvard.edu/abs/2021MNRAS.508.1788A}
}

@INPROCEEDINGS{Bojicic2017,
       author = {{Boji{\v{c}}i{\'c}}, Ivan S. and {Parker}, Quentin A. and {Frew}, David J.},
        title = "{The Hong Kong/AAO/Strasbourg H{\ensuremath{\alpha}} (HASH) Planetary Nebula Database}",
    booktitle = {Planetary Nebulae: Multi-Wavelength Probes of Stellar and Galactic Evolution},
         year = 2017,
       editor = {{Liu}, X. and {Stanghellini}, L. and {Karakas}, A.},
       volume = {323},
        month = oct,
        pages = {327-328},
          doi = {10.1017/S1743921317003234},
archivePrefix = {arXiv},
       eprint = {1612.01521},
 primaryClass = {astro-ph.IM},
       adsurl = {https://ui.adsabs.harvard.edu/abs/2017IAUS..323..327B}
}

@INPROCEEDINGS{Bond1979,
       author = {{Bond}, H.~E.},
        title = "{A Spectroscopic Survey of Cataclysmic Variables}",
    booktitle = {IAU Colloquium 53: White Dwarfs and Variable Degenerate Stars},
         year = 1979,
       editor = {{van Horn}, H.~M. and {Weidemann}, V. and {Savedoff}, M.~P.},
        month = jan,
        pages = {495},
       adsurl = {https://ui.adsabs.harvard.edu/abs/1979wdvd.coll..495B}
}

@ARTICLE{BondPaperI2023,
       author = {{Bond}, Howard E. and {Werner}, Klaus and {Jacoby}, George H. and {Zeimann}, Gregory R.},
        title = "{Spectroscopic survey of faint planetary-nebula nuclei - I. Six new 'O VI' central stars}",
      journal = {\mnras},
         year = 2023,
        month = may,
       volume = {521},
       number = {1},
        pages = {668-676},
          doi = {10.1093/mnras/stad524},
archivePrefix = {arXiv},
       eprint = {2302.07158},
 primaryClass = {astro-ph.SR},
       adsurl = {https://ui.adsabs.harvard.edu/abs/2023MNRAS.521..668B}
}

@ARTICLE{BondV341Ara2018,
       author = {{Bond}, Howard E. and {Miszalski}, Brent},
        title = "{Spectroscopy of V341 Arae: A Nearby Nova-like Variable Inside a Bow Shock Nebula}",
      journal = {\pasp},
         year = 2018,
        month = sep,
       volume = {130},
       number = {991},
        pages = {094201},
          doi = {10.1088/1538-3873/aace3e},
archivePrefix = {arXiv},
       eprint = {1805.11682},
 primaryClass = {astro-ph.SR},
       adsurl = {https://ui.adsabs.harvard.edu/abs/2018PASP..130i4201B}
}

@ARTICLE{BondASASSN2020,
       author = {{Bond}, Howard E.},
        title = "{Bow-Shock Nebula Associated with Novalike Variable ASASSN-V J205457.73+515731.9}",
      journal = {The Astronomer's Telegram},
         year = 2020,
        month = jun,
       volume = {13825},
        pages = {1},
       adsurl = {https://ui.adsabs.harvard.edu/abs/2020ATel13825....1B}
}

@ARTICLE{BondAbell572024,
       author = {{Bond}, Howard E. and {Chaturvedi}, Akshat S. and {Ciardullo}, Robin and {Werner}, Klaus and {Zeimann}, Gregory R. and {Siegel}, Michael H.},
        title = "{Spectroscopic Survey of Faint Planetary-nebula Nuclei. V. The EGB 6-type Central Star of Abell 57}",
      journal = {\apj},
         year = 2024,
        month = aug,
       volume = {970},
       number = {2},
          eid = {164},
        pages = {164},
          doi = {10.3847/1538-4357/ad4f84},
archivePrefix = {arXiv},
       eprint = {2405.11087},
 primaryClass = {astro-ph.SR},
       adsurl = {https://ui.adsabs.harvard.edu/abs/2024ApJ...970..164B}
}

@ARTICLE{BondSYCnc2024,
       author = {{Bond}, Howard E. and {Carter}, Calvin and {Elmore}, David F. and {Goodhew}, Peter and {Patchick}, Dana and {Talbot}, Jonathan},
        title = "{Discovery of a Bow-shock Nebula Around the Z Cam-type Cataclysmic Variable SY Cancri}",
      journal = {\aj},
         year = 2024,
        month = dec,
       volume = {168},
       number = {6},
          eid = {249},
        pages = {249},
          doi = {10.3847/1538-3881/ad7a71},
archivePrefix = {arXiv},
       eprint = {2409.06835},
 primaryClass = {astro-ph.SR},
       adsurl = {https://ui.adsabs.harvard.edu/abs/2024AJ....168..249B},
         note = {(Paper I)}
}

@ARTICLE{BondLSPeg2025,
       author = {{Bond}, Howard E. and {Carter}, Calvin and {Coles}, Eric and {Goodhew}, Peter and {Patchick}, Dana and {Talbot}, Jonathan and {Zeimann}, Gregory R.},
        title = "{Two More Bow Shocks and Off-center H{\ensuremath{\alpha}} Nebulae Associated with Nova-like Cataclysmic Variables}",
      journal = {\aj},
         year = 2025,
        month = aug,
       volume = {170},
       number = {2},
          eid = {78},
        pages = {78},
          doi = {10.3847/1538-3881/add88d},
archivePrefix = {arXiv},
       eprint = {2505.02760},
 primaryClass = {astro-ph.SR},
       adsurl = {https://ui.adsabs.harvard.edu/abs/2025AJ....170...78B},
         note = {(Paper II)}
}

@ARTICLE{BondAntlia2025,
       author = {{Bond}, Howard E. and {Patchick}, Dana and {Stern}, Daniel and {Talbot}, Jonathan and {Thorstensen}, John R.},
        title = "{Discovery of Faint Nebulosity around a Z Camelopardalis{\textendash}type Cataclysmic Variable in Antlia: Nova Shell or Ancient Planetary Nebula?}",
      journal = {\aj},
         year = 2025,
        month = sep,
       volume = {170},
       number = {3},
          eid = {137},
        pages = {137},
          doi = {10.3847/1538-3881/ade9a6},
archivePrefix = {arXiv},
       eprint = {2506.11306},
 primaryClass = {astro-ph.SR},
       adsurl = {https://ui.adsabs.harvard.edu/abs/2025AJ....170..137B},
         note = {(Paper III)}
}

@ARTICLE{Bruch2024,
       author = {{Bruch}, Albert},
        title = "{The AH Pictoris Syndrome: Continuous Trains of Stunted Outbursts in Novalike Variables}",
      journal = {\apj},
         year = 2024,
        month = dec,
       volume = {977},
       number = {2},
          eid = {153},
        pages = {153},
          doi = {10.3847/1538-4357/ad8c39},
archivePrefix = {arXiv},
       eprint = {2410.20756},
 primaryClass = {astro-ph.SR},
       adsurl = {https://ui.adsabs.harvard.edu/abs/2024ApJ...977..153B}
}

@ARTICLE{Casares2009,
       author = {{Casares}, J. and {Mart{\'\i}nez-Pais}, I.~G. and {Rodr{\'\i}guez-Gil}, P.},
        title = "{SY Cnc, a case for unstable mass transfer?}",
      journal = {\mnras},
         year = 2009,
        month = nov,
       volume = {399},
       number = {3},
        pages = {1534-1538},
          doi = {10.1111/j.1365-2966.2009.15384.x},
archivePrefix = {arXiv},
       eprint = {0907.4432},
 primaryClass = {astro-ph.SR},
       adsurl = {https://ui.adsabs.harvard.edu/abs/2009MNRAS.399.1534C}
}

@ARTICLE{CastroSegura2021,
       author = {{Castro Segura}, N. and {Knigge}, C. and {Acosta-Pulido}, J.~A. and {Altamirano}, D. and {del Palacio}, S. and {Hernandez Santisteban}, J.~V. and {Pahari}, M. and {Rodriguez-Gil}, P. and {Belardi}, C. and {Buckley}, D.~A.~H. and {Burleigh}, M.~R. and {Childress}, M. and {Fender}, R.~P. and {Hewitt}, D.~M. and {James}, D.~J. and {Kuhn}, R.~B. and {Kuin}, N.~P.~M. and {Pepper}, J. and {Ponomareva}, A.~A. and {Pretorius}, M.~L. and {Rodr{\'\i}guez}, J.~E. and {Stassun}, K.~G. and {Williams}, D.~R.~A. and {Woudt}, P.~A.},
        title = "{Bow shocks, nova shells, disc winds and tilted discs: the nova-like V341 Ara has it all}",
      journal = {\mnras},
         year = 2021,
        month = feb,
       volume = {501},
       number = {2},
        pages = {1951-1969},
          doi = {10.1093/mnras/staa2516},
archivePrefix = {arXiv},
       eprint = {2008.07462},
 primaryClass = {astro-ph.SR},
       adsurl = {https://ui.adsabs.harvard.edu/abs/2021MNRAS.501.1951C}
}

@ARTICLE{Chen2019,
       author = {{Chen}, B. -Q. and {Huang}, Y. and {Yuan}, H. -B. and {Wang}, C. and {Fan}, D. -W. and {Xiang}, M. -S. and {Zhang}, H. -W. and {Tian}, Z. -J. and {Liu}, X. -W.},
        title = "{Three-dimensional interstellar dust reddening maps of the Galactic plane}",
      journal = {\mnras},
         year = 2019,
        month = mar,
       volume = {483},
       number = {4},
        pages = {4277-4289},
          doi = {10.1093/mnras/sty3341},
archivePrefix = {arXiv},
       eprint = {1807.02241},
 primaryClass = {astro-ph.GA},
       adsurl = {https://ui.adsabs.harvard.edu/abs/2019MNRAS.483.4277C}
}

@INPROCEEDINGS{Chonis2016,
       author = {{Chonis}, Taylor S. and {Hill}, Gary J. and {Lee}, Hanshin and {Tuttle}, Sarah E. and {Vattiat}, Brian L. and {Drory}, Niv and {Indahl}, Briana L. and {Peterson}, Trent W. and {Ramsey}, Jason},
        title = "{LRS2: design, assembly, testing, and commissioning of the second-generation low-resolution spectrograph for the Hobby-Eberly Telescope}",
    booktitle = {Ground-based and Airborne Instrumentation for Astronomy VI},
         year = 2016,
       editor = {{Evans}, Christopher J. and {Simard}, Luc and {Takami}, Hideki},
       series = {Society of Photo-Optical Instrumentation Engineers (SPIE) Conference Series},
       volume = {9908},
        month = aug,
          eid = {99084C},
        pages = {99084C},
          doi = {10.1117/12.2232209},
       adsurl = {https://ui.adsabs.harvard.edu/abs/2016SPIE.9908E..4CC}
}

@ARTICLE{Downes1995,
       author = {{Downes}, Ronald and {Hoard}, D.~W. and {Szkody}, Paula and {Wachter}, Stefanie},
        title = "{Spectroscopy of Poorly Studied Cataclysmic Variables}",
      journal = {\aj},
         year = 1995,
        month = oct,
       volume = {110},
        pages = {1824},
          doi = {10.1086/117654},
       adsurl = {https://ui.adsabs.harvard.edu/abs/1995AJ....110.1824D}
}

@ARTICLE{Drew2005,
       author = {{Drew}, Janet E. and {Greimel}, R. and {Irwin}, M.~J. and {Aungwerojwit}, A. and {Barlow}, M.~J. and {Corradi}, R.~L.~M. and {Drake}, J.~J. and {G{\"a}nsicke}, B.~T. and {Groot}, P. and {Hales}, A. and {Hopewell}, E.~C. and {Irwin}, J. and {Knigge}, C. and {Leisy}, P. and {Lennon}, D.~J. and {Mampaso}, A. and {Masheder}, M.~R.~W. and {Matsuura}, M. and {Morales-Rueda}, L. and {Morris}, R.~A.~H. and {Parker}, Q.~A. and {Phillipps}, S. and {Rodriguez-Gil}, P. and {Roelofs}, G. and {Skillen}, I. and {Sokoloski}, J.~L. and {Steeghs}, D. and {Unruh}, Y.~C. and {Viironen}, K. and {Vink}, J.~S. and {Walton}, N.~A. and {Witham}, A. and {Wright}, N. and {Zijlstra}, A.~A. and {Zurita}, A.},
        title = "{The INT Photometric H{\ensuremath{\alpha}} Survey of the Northern Galactic Plane (IPHAS)}",
      journal = {\mnras},
         year = 2005,
        month = sep,
       volume = {362},
       number = {3},
        pages = {753-776},
          doi = {10.1111/j.1365-2966.2005.09330.x},
archivePrefix = {arXiv},
       eprint = {astro-ph/0506726},
 primaryClass = {astro-ph},
       adsurl = {https://ui.adsabs.harvard.edu/abs/2005MNRAS.362..753D}
}

@ARTICLE{Ellis1984,
       author = {{Ellis}, G.~L. and {Grayson}, E.~T. and {Bond}, H.~E.},
        title = "{A search for faint planetary nebulae on Palomar Sky Survey prints.}",
      journal = {\pasp},
         year = 1984,
        month = apr,
       volume = {96},
        pages = {283-286},
          doi = {10.1086/131333},
       adsurl = {https://ui.adsabs.harvard.edu/abs/1984PASP...96..283E}
}

@INPROCEEDINGS{Frew2006,
       author = {{Frew}, David J. and {Madsen}, G.~J. and {Parker}, Q.~A.},
        title = "{A Search for New Emission Nebulae from the SHASSA and VTSS Surveys}",
    booktitle = {Planetary Nebulae in our Galaxy and Beyond},
         year = 2006,
       editor = {{Barlow}, Michael J. and {M{\'e}ndez}, Roberto H.},
       series = {IAU Symposium},
       volume = {234},
        month = jan,
        pages = {395-396},
          doi = {10.1017/S1743921306003413},
       adsurl = {https://ui.adsabs.harvard.edu/abs/2006IAUS..234..395F}
}

@ARTICLE{Gaia2016,
       author = {{Gaia Collaboration} and {Prusti}, T. and {de Bruijne}, J.~H.~J. and {Brown}, A.~G.~A. and {Vallenari}, A. and {Babusiaux}, C. and {Bailer-Jones}, C.~A.~L. and {Bastian}, U. and {Biermann}, M. and {Evans}, D.~W. and {Eyer}, L. and {Jansen}, F. and {Jordi}, C. and {Klioner}, S.~A. and {Lammers}, U. and {Lindegren}, L. and {Luri}, X. and {Mignard}, F. and {Milligan}, D.~J. and {Panem}, C. and {Poinsignon}, V. and {Pourbaix}, D. and {Randich}, S. and {Sarri}, G. and {Sartoretti}, P. and {Siddiqui}, H.~I. and {Soubiran}, C. and {Valette}, V. and {van Leeuwen}, F. and {Walton}, N.~A. and {Aerts}, C. and {Arenou}, F. and {Cropper}, M. and {Drimmel}, R. and {H{\o}g}, E. and {Katz}, D. and {Lattanzi}, M.~G. and {O'Mullane}, W. and {Grebel}, E.~K. and {Holland}, A.~D. and {Huc}, C. and {Passot}, X. and {Bramante}, L. and {Cacciari}, C. and {Casta{\~n}eda}, J. and {Chaoul}, L. and {Cheek}, N. and {De Angeli}, F. and {Fabricius}, C. and {Guerra}, R. and {Hern{\'a}ndez}, J. and {Jean-Antoine-Piccolo}, A. and {Masana}, E. and {Messineo}, R. and {Mowlavi}, N. and {Nienartowicz}, K. and {Ord{\'o}{\~n}ez-Blanco}, D. and {Panuzzo}, P. and {Portell}, J. and {Richards}, P.~J. and {Riello}, M. and {Seabroke}, G.~M. and {Tanga}, P. and {Th{\'e}venin}, F. and {Torra}, J. and {Els}, S.~G. and {Gracia-Abril}, G. and {Comoretto}, G. and {Garcia-Reinaldos}, M. and {Lock}, T. and {Mercier}, E. and {Altmann}, M. and {Andrae}, R. and {Astraatmadja}, T.~L. and {Bellas-Velidis}, I. and {Benson}, K. and {Berthier}, J. and {Blomme}, R. and {Busso}, G. and {Carry}, B. and {Cellino}, A. and {Clementini}, G. and {Cowell}, S. and {Creevey}, O. and {Cuypers}, J. and {Davidson}, M. and {De Ridder}, J. and {de Torres}, A. and {Delchambre}, L. and {Dell'Oro}, A. and {Ducourant}, C. and {Fr{\'e}mat}, Y. and {Garc{\'\i}a-Torres}, M. and {Gosset}, E. and {Halbwachs}, J. -L. and {Hambly}, N.~C. and {Harrison}, D.~L. and {Hauser}, M. and {Hestroffer}, D. and {Hodgkin}, S.~T. and {Huckle}, H.~E. and {Hutton}, A. and {Jasniewicz}, G. and {Jordan}, S. and {Kontizas}, M. and {Korn}, A.~J. and {Lanzafame}, A.~C. and {Manteiga}, M. and {Moitinho}, A. and {Muinonen}, K. and {Osinde}, J. and {Pancino}, E. and {Pauwels}, T. and {Petit}, J. -M. and {Recio-Blanco}, A. and {Robin}, A.~C. and {Sarro}, L.~M. and {Siopis}, C. and {Smith}, M. and {Smith}, K.~W. and {Sozzetti}, A. and {Thuillot}, W. and {van Reeven}, W. and {Viala}, Y. and {Abbas}, U. and {Abreu Aramburu}, A. and {Accart}, S. and {Aguado}, J.~J. and {Allan}, P.~M. and {Allasia}, W. and {Altavilla}, G. and {{\'A}lvarez}, M.~A. and {Alves}, J. and {Anderson}, R.~I. and {Andrei}, A.~H. and {Anglada Varela}, E. and {Antiche}, E. and {Antoja}, T. and {Ant{\'o}n}, S. and {Arcay}, B. and {Atzei}, A. and {Ayache}, L. and {Bach}, N. and {Baker}, S.~G. and {Balaguer-N{\'u}{\~n}ez}, L. and {Barache}, C. and {Barata}, C. and {Barbier}, A. and {Barblan}, F. and {Baroni}, M. and {Barrado y Navascu{\'e}s}, D. and {Barros}, M. and {Barstow}, M.~A. and {Becciani}, U. and {Bellazzini}, M. and {Bellei}, G. and {Bello Garc{\'\i}a}, A. and {Belokurov}, V. and {Bendjoya}, P. and {Berihuete}, A. and {Bianchi}, L. and {Bienaym{\'e}}, O. and {Billebaud}, F. and {Blagorodnova}, N. and {Blanco-Cuaresma}, S. and {Boch}, T. and {Bombrun}, A. and {Borrachero}, R. and {Bouquillon}, S. and {Bourda}, G. and {Bouy}, H. and {Bragaglia}, A. and {Breddels}, M.~A. and {Brouillet}, N. and {Br{\"u}semeister}, T. and {Bucciarelli}, B. and {Budnik}, F. and {Burgess}, P. and {Burgon}, R. and {Burlacu}, A. and {Busonero}, D. and {Buzzi}, R. and {Caffau}, E. and {Cambras}, J. and {Campbell}, H. and {Cancelliere}, R. and {Cantat-Gaudin}, T. and {Carlucci}, T. and {Carrasco}, J.~M. and {Castellani}, M. and {Charlot}, P. and {Charnas}, J. and {Charvet}, P. and {Chassat}, F. and {Chiavassa}, A. and {Clotet}, M. and {Cocozza}, G. and {Collins}, R.~S. and {Collins}, P. and {Costigan}, G. and {Crifo}, F. and {Cross}, N.~J.~G. and {Crosta}, M. and {Crowley}, C. and {Dafonte}, C. and {Damerdji}, Y. and {Dapergolas}, A. and {David}, P. and {David}, M. and {De Cat}, P. and {de Felice}, F. and {de Laverny}, P. and {De Luise}, F. and {De March}, R. and {de Martino}, D. and {de Souza}, R. and {Debosscher}, J. and {del Pozo}, E. and {Delbo}, M. and {Delgado}, A. and {Delgado}, H.~E. and {di Marco}, F. and {Di Matteo}, P. and {Diakite}, S. and {Distefano}, E. and {Dolding}, C. and {Dos Anjos}, S. and {Drazinos}, P. and {Dur{\'a}n}, J. and {Dzigan}, Y. and {Ecale}, E. and {Edvardsson}, B. and {Enke}, H. and {Erdmann}, M. and {Escolar}, D. and {Espina}, M. and {Evans}, N.~W. and {Eynard Bontemps}, G. and {Fabre}, C. and {Fabrizio}, M. and {Faigler}, S. and {Falc{\~a}o}, A.~J. and {Farr{\`a}s Casas}, M. and {Faye}, F. and {Federici}, L. and {Fedorets}, G. and {Fern{\'a}ndez-Hern{\'a}ndez}, J. and {Fernique}, P. and {Fienga}, A. and {Figueras}, F. and {Filippi}, F. and {Findeisen}, K. and {Fonti}, A. and {Fouesneau}, M. and {Fraile}, E. and {Fraser}, M. and {Fuchs}, J. and {Furnell}, R. and {Gai}, M. and {Galleti}, S. and {Galluccio}, L. and {Garabato}, D. and {Garc{\'\i}a-Sedano}, F. and {Gar{\'e}}, P. and {Garofalo}, A. and {Garralda}, N. and {Gavras}, P. and {Gerssen}, J. and {Geyer}, R. and {Gilmore}, G. and {Girona}, S. and {Giuffrida}, G. and {Gomes}, M. and {Gonz{\'a}lez-Marcos}, A. and {Gonz{\'a}lez-N{\'u}{\~n}ez}, J. and {Gonz{\'a}lez-Vidal}, J.~J. and {Granvik}, M. and {Guerrier}, A. and {Guillout}, P. and {Guiraud}, J. and {G{\'u}rpide}, A. and {Guti{\'e}rrez-S{\'a}nchez}, R. and {Guy}, L.~P. and {Haigron}, R. and {Hatzidimitriou}, D. and {Haywood}, M. and {Heiter}, U. and {Helmi}, A. and {Hobbs}, D. and {Hofmann}, W. and {Holl}, B. and {Holland}, G. and {Hunt}, J.~A.~S. and {Hypki}, A. and {Icardi}, V. and {Irwin}, M. and {Jevardat de Fombelle}, G. and {Jofr{\'e}}, P. and {Jonker}, P.~G. and {Jorissen}, A. and {Julbe}, F. and {Karampelas}, A. and {Kochoska}, A. and {Kohley}, R. and {Kolenberg}, K. and {Kontizas}, E. and {Koposov}, S.~E. and {Kordopatis}, G. and {Koubsky}, P. and {Kowalczyk}, A. and {Krone-Martins}, A. and {Kudryashova}, M. and {Kull}, I. and {Bachchan}, R.~K. and {Lacoste-Seris}, F. and {Lanza}, A.~F. and {Lavigne}, J. -B. and {Le Poncin-Lafitte}, C. and {Lebreton}, Y. and {Lebzelter}, T. and {Leccia}, S. and {Leclerc}, N. and {Lecoeur-Taibi}, I. and {Lemaitre}, V. and {Lenhardt}, H. and {Leroux}, F. and {Liao}, S. and {Licata}, E. and {Lindstr{\o}m}, H.~E.~P. and {Lister}, T.~A. and {Livanou}, E. and {Lobel}, A. and {L{\"o}ffler}, W. and {L{\'o}pez}, M. and {Lopez-Lozano}, A. and {Lorenz}, D. and {Loureiro}, T. and {MacDonald}, I. and {Magalh{\~a}es Fernandes}, T. and {Managau}, S. and {Mann}, R.~G. and {Mantelet}, G. and {Marchal}, O. and {Marchant}, J.~M. and {Marconi}, M. and {Marie}, J. and {Marinoni}, S. and {Marrese}, P.~M. and {Marschalk{\'o}}, G. and {Marshall}, D.~J. and {Mart{\'\i}n-Fleitas}, J.~M. and {Martino}, M. and {Mary}, N. and {Matijevi{\v{c}}}, G. and {Mazeh}, T. and {McMillan}, P.~J. and {Messina}, S. and {Mestre}, A. and {Michalik}, D. and {Millar}, N.~R. and {Miranda}, B.~M.~H. and {Molina}, D. and {Molinaro}, R. and {Molinaro}, M. and {Moln{\'a}r}, L. and {Moniez}, M. and {Montegriffo}, P. and {Monteiro}, D. and {Mor}, R. and {Mora}, A. and {Morbidelli}, R. and {Morel}, T. and {Morgenthaler}, S. and {Morley}, T. and {Morris}, D. and {Mulone}, A.~F. and {Muraveva}, T. and {Musella}, I. and {Narbonne}, J. and {Nelemans}, G. and {Nicastro}, L. and {Noval}, L. and {Ord{\'e}novic}, C. and {Ordieres-Mer{\'e}}, J. and {Osborne}, P. and {Pagani}, C. and {Pagano}, I. and {Pailler}, F. and {Palacin}, H. and {Palaversa}, L. and {Parsons}, P. and {Paulsen}, T. and {Pecoraro}, M. and {Pedrosa}, R. and {Pentik{\"a}inen}, H. and {Pereira}, J. and {Pichon}, B. and {Piersimoni}, A.~M. and {Pineau}, F. -X. and {Plachy}, E. and {Plum}, G. and {Poujoulet}, E. and {Pr{\v{s}}a}, A. and {Pulone}, L. and {Ragaini}, S. and {Rago}, S. and {Rambaux}, N. and {Ramos-Lerate}, M. and {Ranalli}, P. and {Rauw}, G. and {Read}, A. and {Regibo}, S. and {Renk}, F. and {Reyl{\'e}}, C. and {Ribeiro}, R.~A. and {Rimoldini}, L. and {Ripepi}, V. and {Riva}, A. and {Rixon}, G. and {Roelens}, M. and {Romero-G{\'o}mez}, M. and {Rowell}, N. and {Royer}, F. and {Rudolph}, A. and {Ruiz-Dern}, L. and {Sadowski}, G. and {Sagrist{\`a} Sell{\'e}s}, T. and {Sahlmann}, J. and {Salgado}, J. and {Salguero}, E. and {Sarasso}, M. and {Savietto}, H. and {Schnorhk}, A. and {Schultheis}, M. and {Sciacca}, E. and {Segol}, M. and {Segovia}, J.~C. and {Segransan}, D. and {Serpell}, E. and {Shih}, I. -C. and {Smareglia}, R. and {Smart}, R.~L. and {Smith}, C. and {Solano}, E. and {Solitro}, F. and {Sordo}, R. and {Soria Nieto}, S. and {Souchay}, J. and {Spagna}, A. and {Spoto}, F. and {Stampa}, U. and {Steele}, I.~A. and {Steidelm{\"u}ller}, H. and {Stephenson}, C.~A. and {Stoev}, H. and {Suess}, F.~F. and {S{\"u}veges}, M. and {Surdej}, J. and {Szabados}, L. and {Szegedi-Elek}, E. and {Tapiador}, D. and {Taris}, F. and {Tauran}, G. and {Taylor}, M.~B. and {Teixeira}, R. and {Terrett}, D. and {Tingley}, B. and {Trager}, S.~C. and {Turon}, C. and {Ulla}, A. and {Utrilla}, E. and {Valentini}, G. and {van Elteren}, A. and {Van Hemelryck}, E. and {van Leeuwen}, M. and {Varadi}, M. and {Vecchiato}, A. and {Veljanoski}, J. and {Via}, T. and {Vicente}, D. and {Vogt}, S. and {Voss}, H. and {Votruba}, V. and {Voutsinas}, S. and {Walmsley}, G. and {Weiler}, M. and {Weingrill}, K. and {Werner}, D. and {Wevers}, T. and {Whitehead}, G. and {Wyrzykowski}, {\L}. and {Yoldas}, A. and {{\v{Z}}erjal}, M. and {Zucker}, S. and {Zurbach}, C. and {Zwitter}, T. and {Alecu}, A. and {Allen}, M. and {Allende Prieto}, C. and {Amorim}, A. and {Anglada-Escud{\'e}}, G. and {Arsenijevic}, V. and {Azaz}, S. and {Balm}, P. and {Beck}, M. and {Bernstein}, H. -H. and {Bigot}, L. and {Bijaoui}, A. and {Blasco}, C. and {Bonfigli}, M. and {Bono}, G. and {Boudreault}, S. and {Bressan}, A. and {Brown}, S. and {Brunet}, P. -M. and {Bunclark}, P. and {Buonanno}, R. and {Butkevich}, A.~G. and {Carret}, C. and {Carrion}, C. and {Chemin}, L. and {Ch{\'e}reau}, F. and {Corcione}, L. and {Darmigny}, E. and {de Boer}, K.~S. and {de Teodoro}, P. and {de Zeeuw}, P.~T. and {Delle Luche}, C. and {Domingues}, C.~D. and {Dubath}, P. and {Fodor}, F. and {Fr{\'e}zouls}, B. and {Fries}, A. and {Fustes}, D. and {Fyfe}, D. and {Gallardo}, E. and {Gallegos}, J. and {Gardiol}, D. and {Gebran}, M. and {Gomboc}, A. and {G{\'o}mez}, A. and {Grux}, E. and {Gueguen}, A. and {Heyrovsky}, A. and {Hoar}, J. and {Iannicola}, G. and {Isasi Parache}, Y. and {Janotto}, A. -M. and {Joliet}, E. and {Jonckheere}, A. and {Keil}, R. and {Kim}, D. -W. and {Klagyivik}, P. and {Klar}, J. and {Knude}, J. and {Kochukhov}, O. and {Kolka}, I. and {Kos}, J. and {Kutka}, A. and {Lainey}, V. and {LeBouquin}, D. and {Liu}, C. and {Loreggia}, D. and {Makarov}, V.~V. and {Marseille}, M.~G. and {Martayan}, C. and {Martinez-Rubi}, O. and {Massart}, B. and {Meynadier}, F. and {Mignot}, S. and {Munari}, U. and {Nguyen}, A. -T. and {Nordlander}, T. and {Ocvirk}, P. and {O'Flaherty}, K.~S. and {Olias Sanz}, A. and {Ortiz}, P. and {Osorio}, J. and {Oszkiewicz}, D. and {Ouzounis}, A. and {Palmer}, M. and {Park}, P. and {Pasquato}, E. and {Peltzer}, C. and {Peralta}, J. and {P{\'e}turaud}, F. and {Pieniluoma}, T. and {Pigozzi}, E. and {Poels}, J. and {Prat}, G. and {Prod'homme}, T. and {Raison}, F. and {Rebordao}, J.~M. and {Risquez}, D. and {Rocca-Volmerange}, B. and {Rosen}, S. and {Ruiz-Fuertes}, M.~I. and {Russo}, F. and {Sembay}, S. and {Serraller Vizcaino}, I. and {Short}, A. and {Siebert}, A. and {Silva}, H. and {Sinachopoulos}, D. and {Slezak}, E. and {Soffel}, M. and {Sosnowska}, D. and {Strai{\v{z}}ys}, V. and {ter Linden}, M. and {Terrell}, D. and {Theil}, S. and {Tiede}, C. and {Troisi}, L. and {Tsalmantza}, P. and {Tur}, D. and {Vaccari}, M. and {Vachier}, F. and {Valles}, P. and {Van Hamme}, W. and {Veltz}, L. and {Virtanen}, J. and {Wallut}, J. -M. and {Wichmann}, R. and {Wilkinson}, M.~I. and {Ziaeepour}, H. and {Zschocke}, S.},
        title = "{The Gaia mission}",
      journal = {\aap},
         year = 2016,
        month = nov,
       volume = {595},
          eid = {A1},
        pages = {A1},
          doi = {10.1051/0004-6361/201629272},
archivePrefix = {arXiv},
       eprint = {1609.04153},
 primaryClass = {astro-ph.IM},
       adsurl = {https://ui.adsabs.harvard.edu/abs/2016A&A...595A...1G}
}

@ARTICLE{Gaia2023,
       author = {{Gaia Collaboration} and {Vallenari}, A. and {Brown}, A.~G.~A. and {Prusti}, T. and {de Bruijne}, J.~H.~J. and {Arenou}, F. and {Babusiaux}, C. and {Biermann}, M. and {Creevey}, O.~L. and {Ducourant}, C. and et al.},
        title = "{Gaia Data Release 3. Summary of the content and survey properties}",
      journal = {\aap},
         year = 2023,
        month = jun,
       volume = {674},
          eid = {A1},
        pages = {A1},
          doi = {10.1051/0004-6361/202243940},
archivePrefix = {arXiv},
       eprint = {2208.00211},
 primaryClass = {astro-ph.GA},
       adsurl = {https://ui.adsabs.harvard.edu/abs/2023A&A...674A...1G}
}

@ARTICLE{Hill2021,
       author = {{Hill}, Gary J. and {Lee}, Hanshin and {MacQueen}, Phillip J. and {Kelz}, Andreas and {Drory}, Niv and {Vattiat}, Brian L. and {Good}, John M. and {Ramsey}, Jason and {Kriel}, Herman and {Peterson}, Trent and {DePoy}, D.~L. and {Gebhardt}, Karl and {Marshall}, J.~L. and {Tuttle}, Sarah E. and {Bauer}, Svend M. and {Chonis}, Taylor S. and {Fabricius}, Maximilian H. and {Froning}, Cynthia and {H{\"a}user}, Marco and {Indahl}, Briana L. and {Jahn}, Thomas and {Landriau}, Martin and {Leck}, Ron and {Montesano}, Francesco and {Prochaska}, Travis and {Snigula}, Jan M. and {Zeimann}, Greg and {Bryant}, Randy and {Damm}, George and {Fowler}, J.~R. and {Janowiecki}, Steven and {Martin}, Jerry and {Mrozinski}, Emily and {Odewahn}, Stephen and {Rostopchin}, Sergey and {Shetrone}, Matthew and {Spencer}, Renny and {Mentuch Cooper}, Erin and {Armandroff}, Taft and {Bender}, Ralf and {Dalton}, Gavin and {Hopp}, Ulrich and {Komatsu}, Eiichiro and {Nicklas}, Harald and {Ramsey}, Lawrence W. and {Roth}, Martin M. and {Schneider}, Donald P. and {Sneden}, Chris and {Steinmetz}, Matthias},
        title = "{The HETDEX Instrumentation: Hobby-Eberly Telescope Wide-field Upgrade and VIRUS}",
      journal = {\aj},
         year = 2021,
        month = dec,
       volume = {162},
       number = {6},
          eid = {298},
        pages = {298},
          doi = {10.3847/1538-3881/ac2c02},
archivePrefix = {arXiv},
       eprint = {2110.03843},
 primaryClass = {astro-ph.IM},
       adsurl = {https://ui.adsabs.harvard.edu/abs/2021AJ....162..298H}
}

@ARTICLE{Hoffmeister1957,
       author = {{Hoffmeister}, C.},
        title = "{Umgebungskarten f{\"u}r die in Sonneberg entdeckten ver{\"a}nderlichen Sterne}",
      journal = {Zentralinstitut fuer Astrophysik Sternwarte Sonneberg Mitteilungen ueber Veraenderliche Sterne},
         year = 1957,
        month = jan,
       volume = {1},
        pages = {245-330},
       adsurl = {https://ui.adsabs.harvard.edu/abs/1957MitVS...1..245H}
}

@ARTICLE{Jacoby2010,
       author = {{Jacoby}, George H. and {Kronberger}, Matthias and {Patchick}, Dana and {Teutsch}, Philipp and {Saloranta}, Jaakko and {Howell}, Michael and {Crisp}, Richard and {Riddle}, Dave and {Acker}, Agn{\`e}s and {Frew}, David J. and {Parker}, Quentin A.},
        title = "{Searching for Faint Planetary Nebulae Using the Digital Sky Survey}",
      journal = {\pasa},
         year = 2010,
        month = may,
       volume = {27},
       number = {2},
        pages = {156-165},
          doi = {10.1071/AS09025},
archivePrefix = {arXiv},
       eprint = {0910.0465},
 primaryClass = {astro-ph.SR},
       adsurl = {https://ui.adsabs.harvard.edu/abs/2010PASA...27..156J}
}

@ARTICLE{Kato2019,
       author = {{Kato}, Taichi},
        title = "{Three Z Camelopardalis-type dwarf novae exhibiting IW Andromedae-type phenomenon}",
      journal = {\pasj},
         year = 2019,
        month = jan,
       volume = {71},
       number = {1},
          eid = {20},
        pages = {20},
          doi = {10.1093/pasj/psy138},
archivePrefix = {arXiv},
       eprint = {1811.05038},
 primaryClass = {astro-ph.SR},
       adsurl = {https://ui.adsabs.harvard.edu/abs/2019PASJ...71...20K}
}

@ARTICLE{Kimura2020,
       author = {{Kimura}, Mariko and {Osaki}, Yoji and {Kato}, Taichi and {Mineshige}, Shin},
        title = "{Thermal-viscous instability in tilted accretion disks: A possible application to IW Andromeda-type dwarf novae}",
      journal = {\pasj},
         year = 2020,
        month = apr,
       volume = {72},
       number = {2},
          eid = {22},
        pages = {22},
          doi = {10.1093/pasj/psz144},
archivePrefix = {arXiv},
       eprint = {1912.07217},
 primaryClass = {astro-ph.SR},
       adsurl = {https://ui.adsabs.harvard.edu/abs/2020PASJ...72...22K}
}

@ARTICLE{Kochanek2017,
       author = {{Kochanek}, C.~S. and {Shappee}, B.~J. and {Stanek}, K.~Z. and {Holoien}, T.~W. -S. and {Thompson}, Todd A. and {Prieto}, J.~L. and {Dong}, Subo and {Shields}, J.~V. and {Will}, D. and {Britt}, C. and {Perzanowski}, D. and {Pojma{\'n}ski}, G.},
        title = "{The All-Sky Automated Survey for Supernovae (ASAS-SN) Light Curve Server v1.0}",
      journal = {\pasp},
         year = 2017,
        month = oct,
       volume = {129},
       number = {980},
        pages = {104502},
          doi = {10.1088/1538-3873/aa80d9},
archivePrefix = {arXiv},
       eprint = {1706.07060},
 primaryClass = {astro-ph.SR},
       adsurl = {https://ui.adsabs.harvard.edu/abs/2017PASP..129j4502K}
}

@ARTICLE{LeDu2022,
       author = {{Le D{\^u}}, P. and {Mulato}, L. and {Parker}, Q.~A. and {Petit}, T. and {Ritter}, A. and {Drechsler}, M. and {Strottner}, X. and {Patchick}, D. and {Prestgard}, T. and {Garde}, O. and {Outters}, N. and {Raffaelli}, T.},
        title = "{Amateur PN discoveries and their spectral confirmation: A significant new addition to the Galactic PN inventory}",
      journal = {\aap},
         year = 2022,
        month = oct,
       volume = {666},
          eid = {A152},
        pages = {A152},
          doi = {10.1051/0004-6361/202243393},
       adsurl = {https://ui.adsabs.harvard.edu/abs/2022A&A...666A.152L}
}

@ARTICLE{Meinunger1965,
       author = {{Meinunger}, L.},
        title = "{Zwei Ver{\"a}nderliche vom Typ Z Camelopardalis}",
      journal = {Zentralinstitut fuer Astrophysik Sternwarte Sonneberg Mitteilungen ueber Veraenderliche Sterne},
         year = 1965,
        month = nov,
       volume = {3},
        pages = {110-110},
       adsurl = {https://ui.adsabs.harvard.edu/abs/1965MitVS...3..110M}
}

@ARTICLE{Osaki1996,
       author = {{Osaki}, Yoji},
        title = "{Dwarf-Nova Outbursts}",
      journal = {\pasp},
         year = 1996,
        month = jan,
       volume = {108},
        pages = {39},
          doi = {10.1086/133689},
       adsurl = {https://ui.adsabs.harvard.edu/abs/1996PASP..108...39O}
}

@INPROCEEDINGS{Parker2016,
       author = {{Parker}, Quentin A. and {Boji{\v{c}}i{\'c}}, Ivan S. and {Frew}, David J.},
        title = "{HASH: the Hong Kong/AAO/Strasbourg H{\ensuremath{\alpha}} planetary nebula database}",
    booktitle = {Journal of Physics Conference Series},
         year = 2016,
       series = {Journal of Physics Conference Series},
       volume = {728},
        month = jul,
          eid = {032008},
        pages = {032008},
          doi = {10.1088/1742-6596/728/3/032008},
archivePrefix = {arXiv},
       eprint = {1603.07042},
 primaryClass = {astro-ph.SR},
       adsurl = {https://ui.adsabs.harvard.edu/abs/2016JPhCS.728c2008P}
}

@ARTICLE{Patterson1996,
       author = {{Patterson}, Joseph and {Patino}, Rocio and {Thorstensen}, J.~R. and {Harvey}, David and {Skillman}, Davis R. and {Ringwald}, F.~A.},
        title = "{Periods and Quasiperiods in the Cataclysmic Variable BZ Camelopardalis}",
      journal = {\aj},
         year = 1996,
        month = jun,
       volume = {111},
        pages = {2422},
          doi = {10.1086/117976},
       adsurl = {https://ui.adsabs.harvard.edu/abs/1996AJ....111.2422P}
}

@ARTICLE{Reindl2024,
       author = {{Reindl}, Nicole and {Bond}, Howard E. and {Werner}, Klaus and {Zeimann}, Gregory R.},
        title = "{Spectroscopic survey of faint planetary-nebula nuclei: VI. Seventeen hydrogen-rich central stars}",
      journal = {\aap},
         year = 2024,
        month = oct,
       volume = {690},
          eid = {A366},
        pages = {A366},
          doi = {10.1051/0004-6361/202451591},
archivePrefix = {arXiv},
       eprint = {2408.01411},
 primaryClass = {astro-ph.SR},
       adsurl = {https://ui.adsabs.harvard.edu/abs/2024A&A...690A.366R}
}

@BOOK{Sion2023,
       author = {{Sion}, Edward M.},
        title = "{Accreting White Dwarfs; From exoplanetary probes to classical novae and Type 1a supernovae}",
         year = 2023,
          doi = {10.1088/2514-3433/ac930c},
       adsurl = {https://ui.adsabs.harvard.edu/abs/2023awdf.book.....S}
}

\end{document}